\documentclass[aps,showpacs,twocolumn,floats]{revtex4}
\usepackage{amssymb}
\usepackage{graphicx}
\begin{document}
\def\v#1{{\bf #1}}
\newcommand{\up}{\uparrow}
\newcommand{\dn}{\downarrow}
\newcommand{\la}{\langle}
\newcommand{\ra}{\rangle}
\newcommand{\lc}{\lowercase}
\newcommand{\zn}{ZnGa$_2$O$_4\,\,$}
\title{Ferromagnetism in the Fe-substituted spinel semiconductor ZnGa$_2$O$_4$}

\author{T. Maitra and Roser Valent{\'\i}\\
\vspace{0.7cm}
}

\affiliation{Institut f{\"u}r Theoretische Physik, J. W. Goethe Universit{\"a}t, 
Max-von-Laue-Strasse 1, 60438 Frankfurt am Main, Germany}

\bibliographystyle{apsrev}

\pacs{75.30.Et, 75.50.Pp}

\begin{abstract}
Motivated by the recent experimental observation of long range
ferromagnetic order at a relatively high temperature of 200K in the Fe-doped
ZnGa$_2$O$_4$ semiconducting spinel, we propose a possible mechanism for
the observed ferromagnetism in this system. We show, supported by band structure calculations,
 how a model similar to
the double exchange model can be written down for this system and calculate
the ground state phase diagram for the two cases where Fe is doped either at
the tetrahedral position or at the octahedral position.  
We find that in both cases such a
model can account for a stable ferromagnetic phase in a wide range of
parameter space. We also argue that in the limit of high
Fe$^{2+}$ concentration at the tetrahedral positions a description in terms of 
a two band model is essential.
The two $e_g$ orbitals and the hopping between them play a crucial role 
in stabilizing the ferromagnetic phase in this limit. The case when Fe 
is doped simultaneously at both the tetrahedral and
the octahedral position is also discussed.

\end{abstract} 

\maketitle

\section{Introduction}
 Diluted magnetic semiconductors (DMS) are recently being intensively
 studied in connection with their possible application for spintronic
 devices\cite{Ohno_98}.
Special attention has been devoted to the III-V semiconductors\cite{Ohno_98,Dietl_00} which
develop 
long range ferromagnetic order with Curie temperatures of about 100K
upon doping with a low concentration of magnetic 
impurities like Mn.
Since spintronic applications would become widely accessible if the
ferromagnetism is achieved at room temperature, there is a continuous
search for new materials with high Curie temperatures.

In a very recent experiment\cite{risbud}, Risbud and coauthors
tried to dope Fe into \zn by preparing a solid solution [\zn]$_{1-x}$[Fe$_3$O$_4$]$_x$ of
\zn and Fe$_3$O$_4$ with x=0.05,0.10 and 0.15.
Long-range magnetic order was 
observed in all the three samples with Curie temperatures up to 200K
 as well as ferromagnetic hysteresis of the magnetization at low
 temperatures. Interestingly, the saturation magnetic moment which should be 
 4$\mu_B$ per Fe$_3$O$_4$ unit, is about  
 1$\mu_B$ instead. This has been interpreted \cite{risbud} as an indication
of certain fraction
of Fe not contributing to the ferromagnetic long range order (LRO).  
These authors have also performed Mossbauer experiments in order to ascertain 
the oxidation states of Fe in the host semiconductor ZnGa$_2$O$_4$.
 For a $x=0.15$ doped sample
they observed the presence of only Fe$^{3+}$ states with 
some of them
displaying a paramagnetic signal (doublet) and the rest showing magnetically 
ordered Fe$^{3+}$. 
We note 
at this point that in the context of Fe$_3$O$_4$, it has already been 
argued that above the Verwey transition \cite{Szotek_03} 
an  Fe$^{2+}$ cation can be viewed as an Fe$^{3+}$ ion plus a 
delocalized electron. We deliberate further on this point in section VII.  

Motivated by these observations and the difference between this system
and the III-V semiconductors, we investigate in what follows the 
underlying mechanism of the long range ordering of Fe ions in doped \zn considering  
various possible limits of the problem in terms of an effective model which
is based on the band structure calculation of the system. The band structure
calculation, which will be discussed in the next section, gives the information
about the position and nature of the Fe bands, their hybridization with
the bands of the parent compound as well as the active
bands at the Fermi level which will help
us to construct an effective Hamiltonian for the system. The ground state
magnetic phase diagrams of this effective Hamiltonian are then calculated using the parameters derived from
the band structure calculation.

The host semiconductor \zn has a spinel crystal structure AB$_2$O$_4$ with two 
cation sites: Zn$^{2+}$ (A) in a tetrahedral co-ordination and   
Ga$^{3+}$ (B) in an octahedral co-ordination of oxygens. 
Fe$_3$O$_4$ has, on the other hand, an inverse spinel structure with a chemical
 composition Fe$^{3+}_A$[Fe$^{2+}$,Fe$^{3+}$]$_B$O$_4^{-2}$. When Fe is 
 substituted in 
\zn via the solid solution [\zn]$_{1-x}$[Fe$_3$O$_4$]$_x$, it can either replace Zn in the 
tetrahedral position or  Ga in the octahedral position or both. 
 Here we will consider the following two cases:  i) all substituted Fe 
are in
 tetrahedral positions and ii)  all substituted Fe are in octahedral
 positions. 
We also assume the most general case, 
 namely, that Fe can have both Fe$^{3+}$ and Fe$^{2+}$ oxidation states 
 irrespective of whether it is in tetrahedral or octahedral position 
and that it is always in a high-spin state with spin=5/2 and spin=2 
respectively\cite{comment}. 
We will briefly outline the case where Fe ions are in both tetrahedral
and octahedral positions at the end. 

The paper is organized as follows. In Sec. II we discuss  
our band structure calculations and present the density of states for the 
case of Fe 
doped into  tetrahedral positions in \zn. Out of these results we get insight
about the active orbitals at the Fermi level which are relevant for our model 
Hamiltonian. In Sec. III we investigate three 
different limits of the case with Fe occupying tetrahedral
positions and motivate a model similar to double 
exchange for this system. In Sec. IV we present the model 
Hamiltonian and calculate the  magnetic phase diagram.  Sec. V deals 
with the effect of the Coulomb correlation on the phase diagram results.
 In Sec. VI we study the case
of Fe doping in the octahedral positions and finally in the last section
 we discuss our results and 
make a comparison with the experimental observations.

\section{Bandstructure Calculations}
\zn is a direct band gap semiconductor with an energy gap of about 4.1eV
\cite{sampath1,Itoh_91}. Previous  
band structure calculations \cite{sampath} for \zn showed that the valence
states
 right below  the Fermi level are mostly of oxygen  character with the
 contribution of Zn and Ga being very small.  In order to investigate
the effect of the doping of Fe in the band structure of
 \zn, we considered modified unit cells of ZnGa$_2$O$_4$ with different
Fe content,  i.e. Fe substituting Zn in
 tetrahedral sites in an Fe:Zn ratio 1:2 (50$\%$ doping of Fe) and Fe substituting Ga in octahedral
 sites in an Fe:Ga ratio 1:4 (25$\%$ doping of Fe). 
 These percentage doping of Fe in Zn (or A) sites and in Ga (or B) sites are close to the 
 experimental ones with $x=0.15$ in [\zn]$_{1-x}$[Fe$_3$O$_4$]$_x$ formula unit which are 
 45$\%$ when Fe substitutes Zn only and 22.5 $\%$ when Fe substitutes Ga only.
Here we will illustrate the tetrahedral substitution.

 \zn crystallizes in a normal spinel structure 
with space group $Fd\bar{3}m$ and the primitive (rhombohedral) unit cell contains two
formula units. In order to 
substitute Fe in one of the Zn positions, one has to make the two Zn
positions in the unit cell nonequivalent. One maximal subgroup of $Fd\bar{3}m$ 
which allows for
this substitution  is $F\bar{4}3m$. In this new space group
 we have in addition to Zn, Fe and Ga, two nonequivalent
 oxygen positions O$_1$ and O$_2$. Fe and Zn  are surrounded by O$_1$ and O$_2$ ions respectively
 in a tetrahedral environment, whereas Ga is in an octahedral surrounding with three O$_1$
 and three O$_2$ atoms
\cite{ITA}.

\begin{figure}
\includegraphics[clip=true,width=90mm]{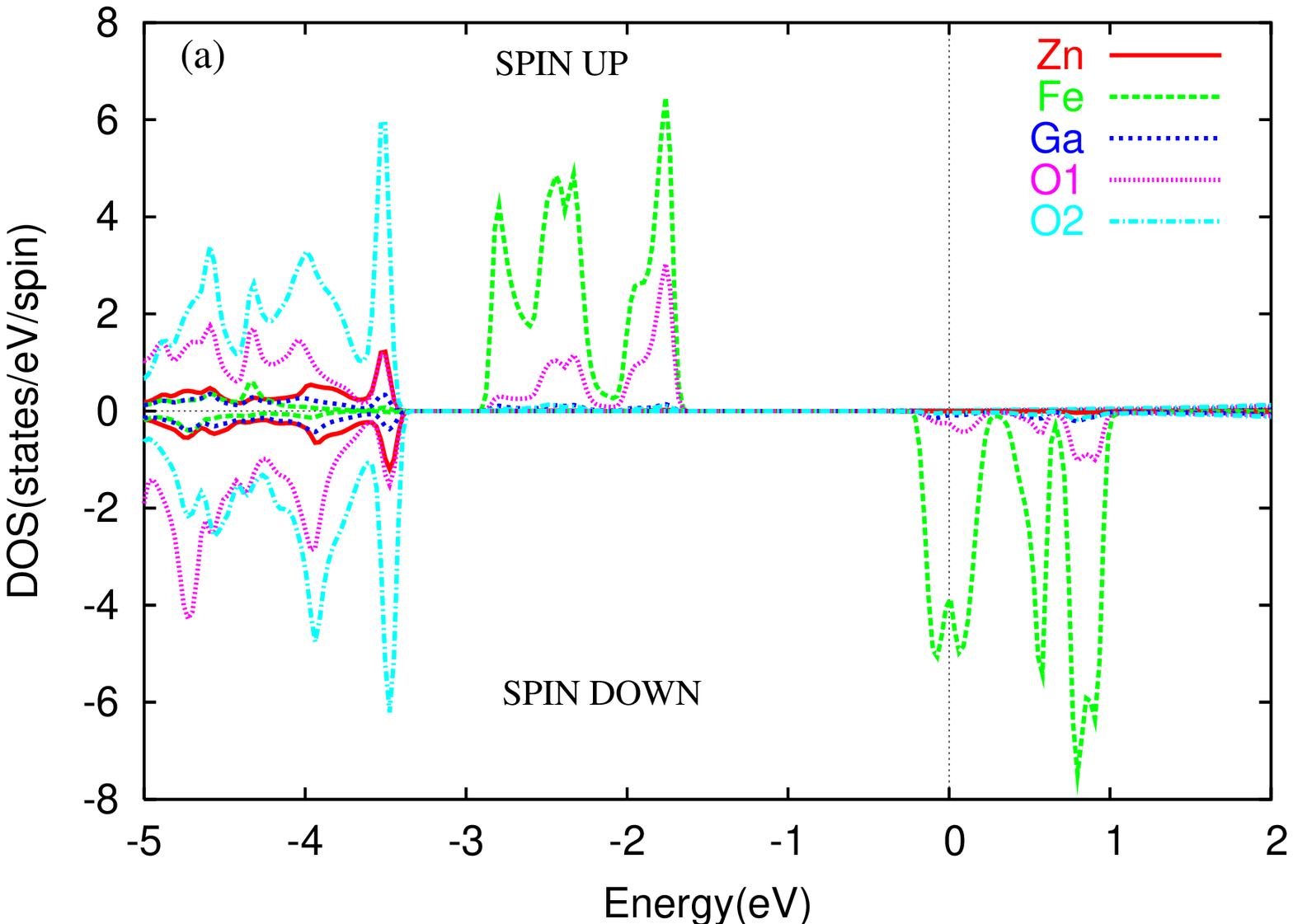}
\includegraphics[clip=true,width=90mm]{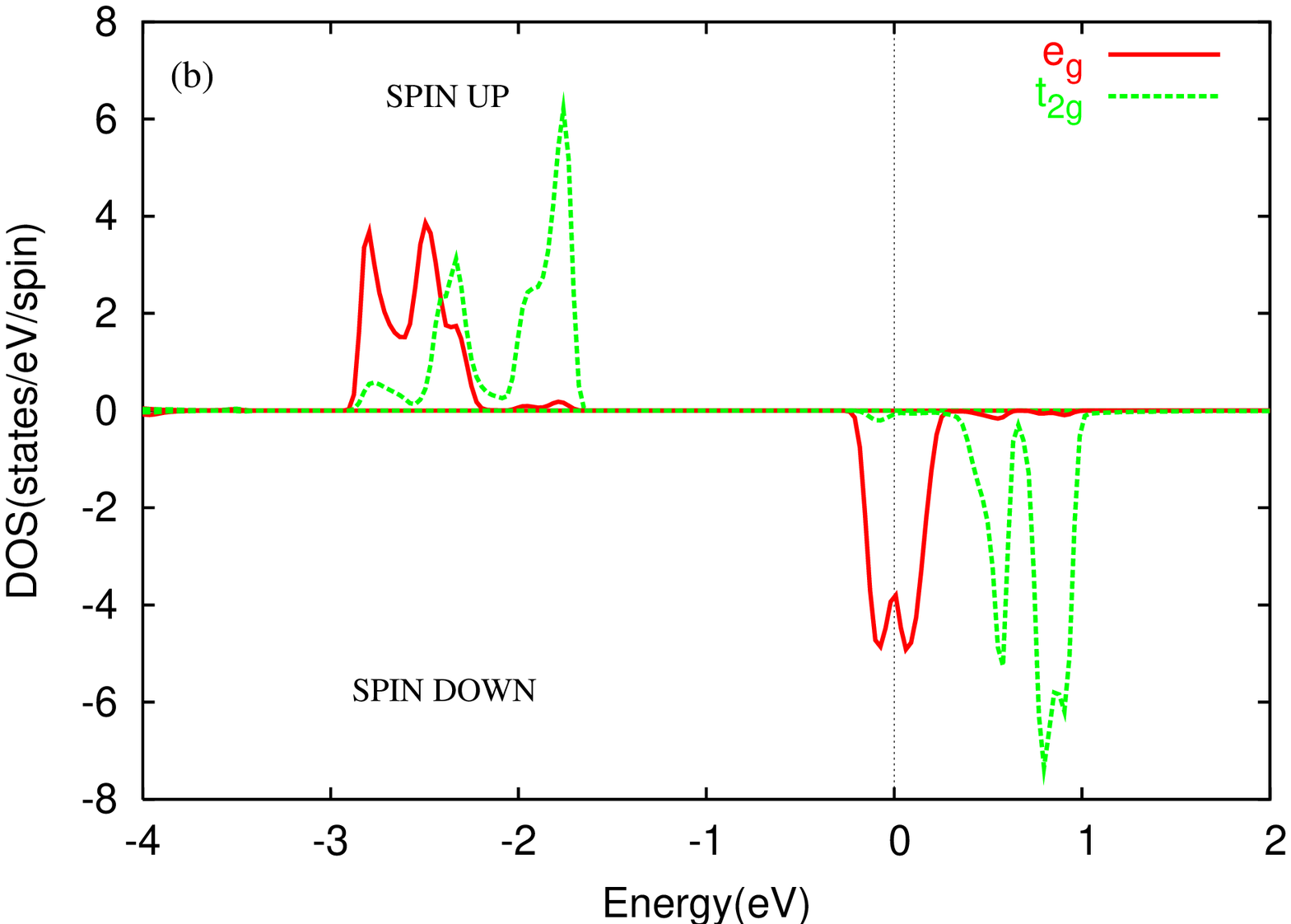}
\caption{
(Color) Majority (up)  and minority  (down)
 spin density of states for Fe doped into the tetrahedral
Zn position in \zn. (a) shows the total density of states and 
(b) the partial Fe-$d$ density of states as explained in the text. Here O1
and O2 denote two nonequivalent oxygen atoms of the modified unit cell.
}
\label{figure1}
\end{figure}

We have performed {\it ab initio} density functional theory  
calculations for Fe doped \zn in the $F\bar{4}3m$ symmetry
 within the local spin density approximation (LSDA)
using the linearized augmented plane waves (LAPW) as basis set \cite{wien2k}.
  In Fig. \ref{figure1} we present the spin polarized density of states (DOS) for Fe 
doped \zn  where Fe substitutes 50$\%$ of tetrahedral Zn atoms in the unit cell as
described above. Total density
of states of majority (up) and minority (down) spin states for all the atoms 
are given in Fig. \ref{figure1}(a) while in  Fig. \ref{figure1}(b) we show the partial density of states
for Fe $d-$states projected into t$_{2g}$ and e$_g$ symmetries in both spin 
directions.
 We observe that the Fe $d$-states 
appear to be mainly located in the band gap of the host semiconductor
 together with a non-negligible contribution of O$_1$ $p$-states  while  
Zn, Ga and O$_2$ contributions are well down into the valence band
and are negligible at the Fermi level (Fig. \ref{figure1}(a)).
 The Fe DOS
  shows the expected $e_g$-$t_{2g}$ splitting for a transition metal 
ion in a tetrahedral crystal field.
The majority spin (spin up) $e_g$  and $t_{2g}$  states are completely
filled and appear far below the Fermi level whereas the minority spin
(spin down) $e_g$
states are partially filled and are at the Fermi level (Fig. \ref{figure1}(b)).
It is important to note that the spin exchange
splitting ($\sim$ 2.6eV) is much larger than the crystal field splitting
($\sim$ 0.5eV) in this system.
These observations of the band structure calculation are crucial to 
build up an effective model for this system, as we will see below.

%%%%%%%%%%%%%%%%%%%%%%%%%%%%%%%%%%%%%%%%%%%%%%%%%%%%%%%%%%%%%%%%%%%%%%%
%%%%%%%%%%%%%%%%%%%%%%%%%%%%%%%%%%%%%%%%%%%%%%%%%%%%%%%%%%%%%%%%%%%%%%%

\section{Model : F\lc{e} in tetrahedral position}

%%%%%%%%%%%%%%%%%%%%%%%%%%%%%%%%%%%%%%%%%%%%%%%%%%%%%%%%%%%%%%%%%%%%%%%
\begin{figure}
\includegraphics[clip=true,width=80mm]{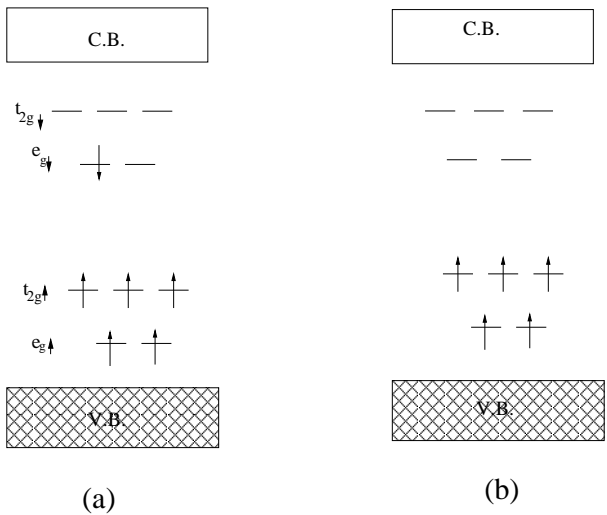}
\caption{Schematic energy level diagram showing the 3$d$ states of doped Fe in the band 
gap of \zn semiconductor.  (a) Fe$^{2+}$ and (b) Fe$^{3+}$ in a tetrahedral
crystal field. V.B. and C.B. denote valence band and conduction band of
 the semiconductor respectively.
}
\label{figure2}
\end{figure}
%%%%%%%%%%%%%%%%%%%%%%%%%%%%%%%%%%%%%%%%%%%%%%%%%%%%%%%%%%%%%%%%%%%%%%%

Based on the electronic structure calculations des\-cribed above for the 
doped system, we now motivate a possible mechanism for the ferromagnetic 
LRO observed in Fe doped \zn. We note at this point that
the Fe concentration in this system is {\it not} very low as
 for x=0.15 in the [\zn]$_{1-x}$[Fe$_3$O$_4$]$_x$ formula unit 
 the Fe:Zn ratio 
 calculated from the nominal composition is 1:1.9 (3x:1-x) or close to 1:2 though
 Risbud {\it et.
al.} reported to have observed a Fe:Zn ratio of 1:3 \cite{risbud}.  
Hence the system is not likely to be a candidate for description in terms of a 
Kondo impurity model \cite{priya}.
It is rather much more reasonable to assume that the
electrons  hop between doped sites via mainly the oxygen orbitals
in the same way as in the double exchange mechanism in the manganites
\cite{dagrev}. We would like to note here that in the case of manganites each Mn 
ion is surrounded by oxygen atoms in
an octahedral environment and these octahedra are corner shared. The double
exchange mechanism then operates through a path of the type Mn-O-Mn giving rise
to a ferromagnetic order among the Mn spins.
 In the present system we observe that for the doping considered in our band structure 
 calculation there exist paths of the type 
 Fe-O$_1$-O$_1$-Fe where O$_1$ is one of the two nonequivalent oxygens which 
 surrounds Fe in a tetrahedral environment as described in the previous section. 
 This assumption of electron transport via oxygen orbitals
is  supported by our band structure
calculations where we see that there exists a small but finite hybridization 
between the doped
Fe and O$_1$ at the Fermi level (see Fig. \ref{figure1}(a)). Note that the contribution of 
 Zn, Ga and  O$_2$ is negligible at the Fermi level and hence are very unlikely
to take part in the electronic transport.

A  double exchange like mechanism mediating ferromagnetism has 
been proposed  
and is being seriously investigated lately for various
diluted magnetic semiconductors\cite{katayama,dietl,pearton}.
First principles calculations have also been seen to support such 
a mechanism for
ferromagnetic order in some of these systems like ZnO based DMS, Ga(Mn)As 
\cite{katayama,sanvito}. 
Indeed, without such a long range transport of electrons,
the ferromagnetic LRO (Long Range Order) observed in these systems would be difficult to account 
for, a view shared in other theoretical analyses\cite{millis} of LRO in DMS. We would 
like to emphasize though that the Fe doped \zn that we are considering here
is not strictly a DMS as the doping level is quite high.

Let us consider first the case where all the doped 
Fe are in  tetrahedral positions. In Fig. 2  we draw a schematic energy level diagram 
of 
Fe 3$d$ orbitals together with the valence and conduction band of the host semiconductor.
We discuss in the following three limits of the problem, 
(i) all or most of the Fe are  
Fe$^{3+}$, (ii) Fe is in both Fe$^{2+}$ and Fe$^{3+}$ oxidation states and (iii) the limit
of high Fe$^{2+}$ concentration.

\subsection{High Fe$^{3+}$ concentration limit}

In the situation when neighbouring doped tetrahedral
sites are all  
Fe$^{3+}$, there are 5 electrons of the same spin in each site due
to Hund's rule. If the spins at these neighbouring doped sites are 
ferromagnetically aligned then
hopping of an electron between  sites  is blocked by 
the Pauli principle. 
If instead, they are antiferromagnetically aligned the system gains
superexchange energy by a virtual process of electron transfer between the 
Fe ions. Therefore,  when all Fe ions are in a Fe$^{3+}$ state, an 
antiferromagnetic (AFM) alignment of spins is energetically preferred. 

\subsection{  Mixed Fe$^{2+}$  and  Fe$^{3+}$ }   

%%%%%%%%%%%%%%%%%%%%%%%%%%%%%%%%%%%%%%%%%%%%%%%%%%%%%%%%%%%%%%%%%%%%%%%
\begin{figure}
\includegraphics[clip=true,width=70mm]{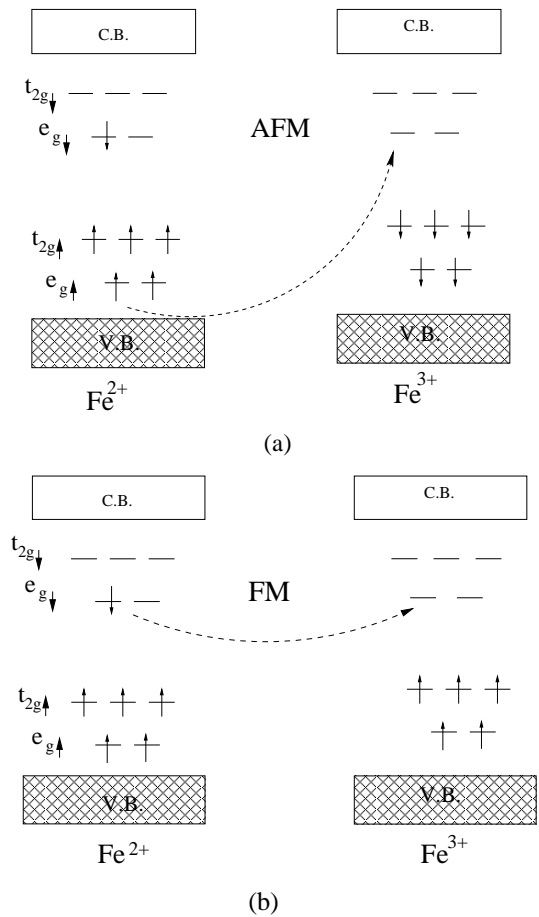}
\caption{
Electron hopping between Fe$^{2+}$ and  Fe$^{3+}$ at
 neighbouring tetrahedral
doped sites with (a) antiferromagnetic alignment of spins (b) ferromagnetic
alignment of spins.
}
\label{figure3}
\end{figure}
%%%%%%%%%%%%%%%%%%%%%%%%%%%%%%%%%%%%%%%%%%%%%%%%%%%%%%%%%%%%%%%%%%%%%%%
Let us examine now the case where both Fe$^{2+}$ and Fe$^{3+}$ are present in 
 neighbouring sites (see Fig.\ \ref{figure3}). If the 5  
electrons in Fe$^{2+}$ (all aligned) are in an antiferromagnetic configuration 
with the spins of the neighbouring Fe$^{3+}$ (Fig.\ \ref{figure3}(a)), then
an electron in Fe$^{2+}$ has to pay an amount of energy equal to 
the Hund's coupling 
($J_H$) in order to hop from one Fe$^{2+}$ site to
a neighbouring Fe$^{3+}$ site. In the limit of large $J_H$ this is practically 
forbidden and the system will try to gain the superexchange energy, 
approximately $\sim$ $\frac{t^2}{J_H},$ where $t$ is the appropriate hopping 
integral between the relevant orbitals \cite{ftn3}. On the contrary, in a  
ferromagnetic arrangement(Fig.\ \ref{figure3}(b)), the minority spin 
electron in Fe$^{2+}$ can move to a neighbouring Fe$^{3+}$ site without paying 
extra energy and the system gains kinetic energy (KE) in this process. 
From the above 
discussion, it is evident that when $J_H$ is large, the system will prefer   
to be in the ferromagnetic state rather than the antiferromagnetic one. 
If $J_H$ is moderate then all these energy scales are comparable 
and the competition between kinetic energy, superexchange energy (SE) and $J_H$ 
will decide the phase boundaries.

It is useful to make a note at this point that the Fe$^{2+}$ ion with 3 
electrons in its $e_g$ 
orbitals is likely to be Jahn-Teller (JT) active, i.e. the doubly-occupied 
orbital 
stabilizes over the singly occupied one, whereas Fe$^{3+}$ with 
two electrons in the two $e_g$ orbitals is not JT active. In this case one 
can then work with only 
one $e_g$ orbital for Fe$^{2+}$ (the Jahn-Teller stabilized one) with 
electrons hopping through this orbital. The mechanism of magnetic exchange due to electron delocalization
goes through without loss of generality as outlined above (with possible
reduction in the overall $e_g$ bandwidth, which can be scaled away). The 
third $e_g$ electron in the Fe$^{+2}$ site and the corresponding $e_g$ 
orbital can be ignored.  But as we will see in the following, this single 
orbital model is not sufficient to describe the case where all (or most of the)
doped Fe are in a Fe$^{2+}$ state. One has then
to take into account both $e_g$ orbitals and the hopping among them. 

\subsection{ High Fe$^{2+}$  concentration limit}   

Here we consider two neighbouring tetrahedral Fe ions, both in 
Fe$^{2+}$ configuration. Since each of them have 3 electrons in 
their $e_g$ orbitals, one $e_g$ orbital is full and the other has one 
electron. Therefore the
only way electrons can hop is via this half-filled orbital. In this case,
there are several possibilities arising from the rela\-tive values of the
JT stabilization energy ($\Delta$) and the bandwidth ($W$) of the e$_g$ 
bands ($\Delta << J_H$ for the system under consideration). Consider the 
situation when $\Delta > W$. In this case ferromagnetism
is inhibited if the hopping matrix is diagonal. If, however, there exists
off-diagonal hopping $t^{12}$ (which, in general, depends on the orbitals 
involved and the symmetry of the lattice), then ferromagnetism could stabilize 
via
a virtual hopping with a gain of ferromagnetic exchange energy $(t^{12})^{2}/
\Delta$. This FM phase is not driven by KE as in the double exchange 
mechanism. However, there is also a competing AFM phase that gains 
superexchange energy of order $t^{2}/(\Delta+J_{H})$. 

In the  limit  $\Delta < W$, the FM state is driven by the double exchange mechanism as 
in the Fe$^{2+}$ - Fe$^{3+}$ mixed configuration. The FM state is stabilized 
by the KE of the e$_g$ electrons since the superexchange energies are less than the KE. The underlying ground state, though, 
will be different when only diagonal hopping is allowed. In this case,
in order to gain the KE, the Fe$^{2+}$ ions will remain in a cooperative, 
staggered JT distorted arrangement which costs additional energy which 
depends on $\Delta$  and may not be stable if AFM superexchange energy is larger
\cite{ftn2}. 

%%%%%%%%%%%%%%%%%%%%%%%%%%%%%%%%%%%%%%%%%%%%%%%%%%%%%%%%%%%%%%%%%%%%%%%
\begin{figure}
\includegraphics[clip=true,width=70mm]{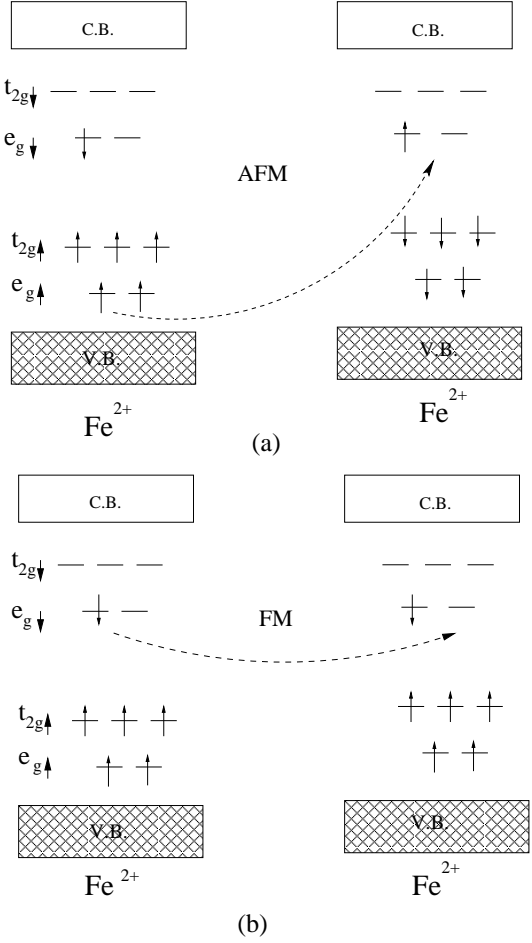}
\caption{Electron hopping in (a) antiferromagnetic phase (b) ferromagnetic
phase when both  neighbouring doped tetrahedral sites are Fe$^{2+}$. 
}
\label{figure4}
\end{figure}
%%%%%%%%%%%%%%%%%%%%%%%%%%%%%%%%%%%%%%%%%%%%%%%%%%%%%%%%%%%%%%%%%%%%%%%

In Fig.\ \ref{figure4} we show the situation arising in the case
$\Delta=0$ with the
$e_g$ orbitals in a cubic environment and nonzero  overlap among them. 
One electron (either up or down) from the doubly occupied $e_g$ orbital 
of Fe$^{2+}$ on one site can always move to the singly occupied $e_g$ orbital 
on the next  Fe$^{2+}$ site. In case of antiferromagnetic alignment
(shown in Fig.\ \ref{figure4}(a)),  
the hopping of an electron from one Fe$^{2+}$ site to the next costs $J_H$ 
amount of energy.  
The superexchange mechanism through virtual hopping is the only energy gain. 
In the ferromagnetic state (Fig.\ \ref{figure4}(b)), however, the initial and 
final 
states are degenerate and the system gains kinetic energy due to resonance. 
It is also evident that the physics for large Fe$^{2+}$ concentration 
is different from the low  Fe$^{2+}$ concentration limit. In the low 
concentration limit, 
an effective single orbital model captures the physical situation well,  while
in the high concentration limit, two bands are crucial for its understanding. 
This situation 
is somewhat reminiscent of the manganites where in the electron-doped side 
(hole concentration $x >$ 0.5) the two $e_g$ orbitals and the hopping 
between them play a crucial role in determining\cite{tm1} the 
competition between the different magnetic phases whereas in the hole-dope 
side (x $<$ 0.5), a model with only the Jahn-Teller stabilized single orbital 
is adequate. 

\section{Hamiltonian}  

Based on the above discussion we write down the following effective 
Hamiltonian for Fe-doped \zn  
assuming a cubic environment

\begin{eqnarray}
H&=&H_0+H_{int}\nonumber\\
H_0&=&\sum_{<ij>,\sigma,\alpha,\beta} t_{ij}^{\alpha\beta}c_{i\alpha\sigma}^{\dagger}
c_{j\beta\sigma}-J_H\sum_iS_i . \sigma_i\nonumber\\
&&\hspace{2.cm}+J_{AF}\sum_{<ij>}S_i . S_j\nonumber\\ 
H_{int}&=&U\sum_{i\alpha}n_{i\alpha\uparrow}n_{i\alpha\downarrow}+U^{\prime}
\sum_{i,\alpha\neq\beta} n_{i\alpha}n_{i\beta}\nonumber\\
\label{H0}
\end{eqnarray}

Here we treat the t$_{2g}$ electrons as localized and $e_g$ as itinerant
because the density of states (Fig. 1) clearly shows that the $e_g$ down 
band is at the Fermi level, whereas the fully filled $t_{2g}$ up band is 
well below. Electronic transport, therefore, involves the electrons in the
$e_g$ band primarily. The $t_{2g}$ bands are well removed from the Fermi
level and the $e_g$-t$_{2g}$ overlap is negligibly small\cite{slater}.
The t$_{2g}$ electrons, under these conditions, provide a localized
magnetic background, to which the itinerant $e_g$ electrons are coupled
through the Hund's exchange.

 The first term in $H_0$ describes the kinetic energy with
$t_{ij}^{\alpha\beta}$ being the anisotropic hopping integral between
two $e_g$ orbitals\cite {slater,pwd,kugel}. Here $i, j$ are site indices and
$\alpha, \beta$ =1,2 are $e_g$ orbital indices. The second term is the Hund's 
coupling term 
between the localized t$_{2g}$ spins and the itinerant $e_g$ spins and the 
last term represents the antiferromagnetic superexchange coupling between  
neighbouring t$_{2g}$ spins. The first and second term in $H_{int}$ define
the onsite intra- and inter-orbital Coulomb repulsion with $U$ and $U^\prime$ 
being 
the corresponding interaction strengths.  
In the 
half-filled situation when we have all Fe in Fe$^{3+}$ states the ferromagnetic
phase is blocked by the Pauli principle and an antiferromagnetic phase is favoured
as we discussed above. The third term in $H_0$ 
representing the antiferromagnetic superexchange satisfies this limit.

First we discuss the model without considering the Coulomb correlation terms
given by $H_{int}$ (Eq.\ \ref{H0}). The correlation and their effects will be
discussed later in detail.
We treat the t$_{2g}$ spin subsystem with a magnitude 3/2 semiclassically as 
 it is also the standard practice in the 
case of manganites \cite{dagrev}. We do not, however, make the further 
assumption prevalent in manganite and DMS literature i.e., 
J$_H\to\infty$. Indeed, such an assumption 
would preclude the presence of a Fe$^{+2}$ state. J$_H$ in the foregoing is
treated as a parameter and its value, as gleaned from the spin splitting
observed in the band structure calculations, is typically large 
(about 2.6$eV$). Assuming an uncanted homogeneous ground state, we choose
$S_i=S_0exp{(i{\bf Q.r}_i)}$ where $S_0=3/2$ and ${\bf Q}=(0,0,0)$ for the 
ferromagnetic 
phase and ${\bf Q}=(\pi,\pi,\pi)$ for the antiferromagnetic phase. With 
this choice the first two terms of the Hamiltonian  $H_0$ (1) reduce to 

\begin{eqnarray}
 H_1&=&\sum \epsilon_{k}^{\alpha\beta}c_{k\alpha\sigma}^{\dagger}
 c_{k\beta\sigma}
 -J_HS_0\sum c_{k\alpha\uparrow}^{\dagger}
 c_{k+Q\alpha\uparrow}\nonumber\\
 &&\hspace{1.5cm}+J_HS_0\sum c_{k\alpha\downarrow}^{\dagger}
 c_{k+Q\alpha\downarrow}\
\label{H1}
\end{eqnarray}

\begin{eqnarray}
\epsilon_k^{11}&=&-2t(cosk_x+cosk_y)\nonumber\\
\epsilon_k^{12}&=&\epsilon_k^{21}=-\frac{2}{\sqrt{3}}t(cosk_x-cosk_y)\nonumber\\
\epsilon_k^{22}&=&-\frac{2}{3}t(cosk_x+cosk_y)-\frac{8}{3}tcosk_z
\end{eqnarray}

Here 1 corresponds to $d_{x^2-y^2}$ and 2 to $d_{3z^2-r^2}$ orbital
and $t$ is the magnitude of the hopping integral between two
neighbouring  $d_{x^2-y^2}$ orbitals in the $x,y$ direction.
The superexchange contribution to the Hamiltonian is given by 
\begin{eqnarray}
E_{SE}&=&\frac{J_{AF}S_0^2}{2}(2cos\theta_{xy}+cos\theta_z) 
\end{eqnarray}
where $\theta_{xy}$ and $\theta_{z}$
are the angles between neighbouring spins in the xy plane and 
in the z-direction respectively.  $\theta_{xy}=\theta_{z}=0$ for the ferromagnetic phase
and $\theta_{xy}=\theta_{z}=\pi$ in the antiferromagnetic phase. These two
angles could be different from $\pi$ or 0 in general and allow for canting.  

\begin{figure}
\includegraphics[clip=true,width=80mm]{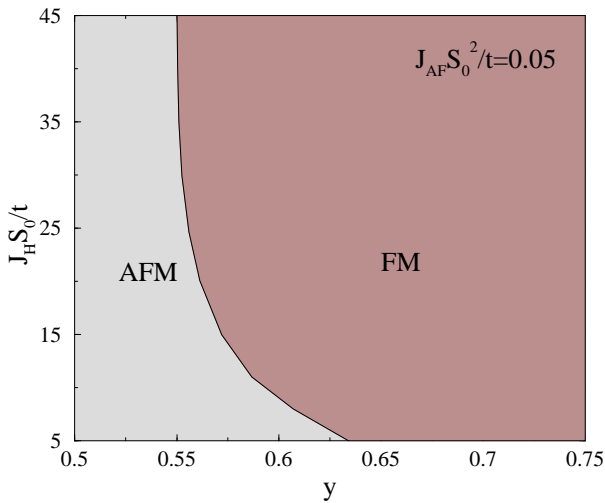}
\caption{Ground state phase diagram of the Hamiltonian ($U=U^\prime=0$) in the
 $y-J_HS_0/t$ plane 
 where $y$ is the $e_g$ electron concentration and with 
$J_{AF}S_0^2/t=0.05$. }
\label{figure5}
\end{figure}

\begin{figure}
\includegraphics[clip=true,width=80mm]{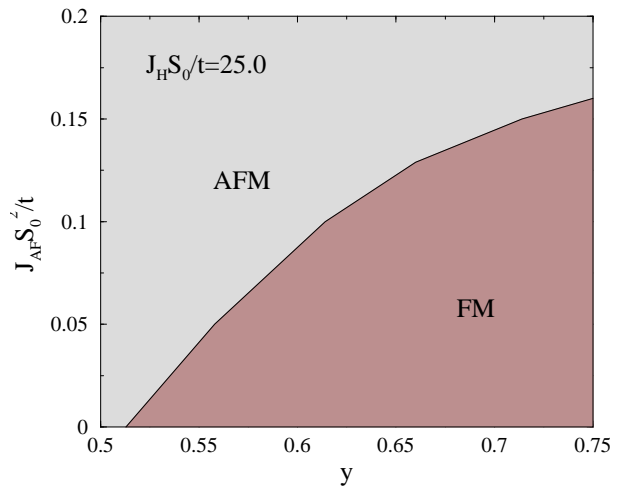}
\caption{Phase diagram for Fe in a tetrahedral environment
in the $y-J_{AF}S_0^2/t$ plane considering both 
$e_g$ orbitals and the hopping among them in the model in the limit $U=U^\prime=0$. 
}
\label{figure6}
\end{figure}

We diagonalize the Hamiltonian (Eq.\ \ref{H1}) at each {\bf k} point on a finite momentum 
grid and calculate the ground state energy for ferromagnetic and 
antiferromagnetic states in their uncanted spin configurations. The magnetic 
structure with minimum ground state energy is determined for each set of 
parameters ($y$, $J_H$ and $J_{AF}$), the two e$_g$ orbitals are taken to be 
degenerate presently ($\Delta=0$). Here $y$ is the $e_g$ electron 
concentration, $y=0.5$ corresponds to the limit where all Fe ions are in 
their Fe$^{3+}$ state and $y=0.75$ corresponds to all Fe in Fe$^{2+}$ state. 
In Fig.\ \ref{figure5} we show the ground state phase diagram in the 
$y-J_HS_0/t$ plane with $J_{AF}S_0^2/t=0.05$ \cite{ftn1} as an illustration.
However, this value of $J_{AF}S_0^2/t$ is 
varied in a wide range to obtain the phase diagram given in Fig. \ref{figure6}.
%%%%%%%%%%%%%%%%%%%%%%%%%%%%%%%%%%%%%%%%%%%%%%%%%%%%%%%%%%%%%%%%%%%%%%%^M
%%%%%%%%%%%%%%%%%%%%%%%%%%%%%%%%%%%%%%%%%%%%%%%%%%%%%%%%%%%%%%%%%%%%%%%^M
In the above figure we see that at y=0.5 where all Fe ions are in their 
Fe$^{3+}$ state the system is antiferromagnetic at all values of
the Hund's coupling $J_H$ as we expected since the ferromagnetic state
is blocked by the Pauli exclusion 
principle in this limit. As we increase the concentration of Fe$^{2+}$, 
the electron concentration increases in the down spin band which can hop 
from site to site and the system gains kinetic energy. Due to the competition, 
modulated by the value of J$_H$, between the kinetic energy which favours 
a FM configuration and superexchange energy which  favours an AFM state, 
a ferromagnetic phase is indeed stabilized over the antiferromagnetic 
one for moderate to high concentration of Fe$^{2+}$.  As the value of $J_H$ is 
increased, the ferromagnetic phase becomes 
broader and at very large $J_H$ the ferromagnetic region becomes almost
independent of $J_H$. 

In Fig.\ \ref{figure6} we present the ground state phase diagram in the $y-J_{AF}S_0/t$ plane
at a typical value of $J_HS_0/t=25.0$ which is again estimated from the 
electronic structure calculation described above. Note that there exists
a wide region in parameter space where the ferromagnetic phase is stabilized.

%%%%%%%%%%%%%%%%%%%%%%%%%%%%%%%%%%%%%%%%%%%%%%%%%%%%%%%%%%%%%%%%%%%%%%%^M
%%%%%%%%%%%%%%%%%%%%%%%%%%%%%%%%%%%%%%%%%%%%%%%%%%%%%%%%%%%%%%%%%%%%%%%^M

In the limit of non-degenerate e$_g$ orbitals, we examine the situation 
for $J_{_H} > \Delta > W$. As argued earlier, there exists the possibility of a 
ferromagnetic phase via double exchange here too in the region of mixed 
Fe$^{+2}$-Fe$^{+3}$ shown in Fig. 7. In this case, the AFM state 
reappears as the Fe$^{+2}$ concentration increases because of reduced effective 
hopping. The phase diagram is symmetric about $y=0.625$ and the regions of 
stability of these phases are nearly independent of $\Delta$ for $\Delta > 
W$ as expected.  
The rather interesting possibilities involving orbital order have not been 
discussed here. The orbital order can be generated by the anisotropic hopping 
as well as the JT distortion. It can also be enhanced by the Coulomb 
correlations\cite{tm1}.  
%%%%%%%%%%%%%%%%%%%%%%%%%%%%%%%%%%%%%%%%%%%%%%%%%%%%%%%%%%%%%%%%%%%%%%%^M
\begin{figure}
\includegraphics[clip=true,width=80mm]{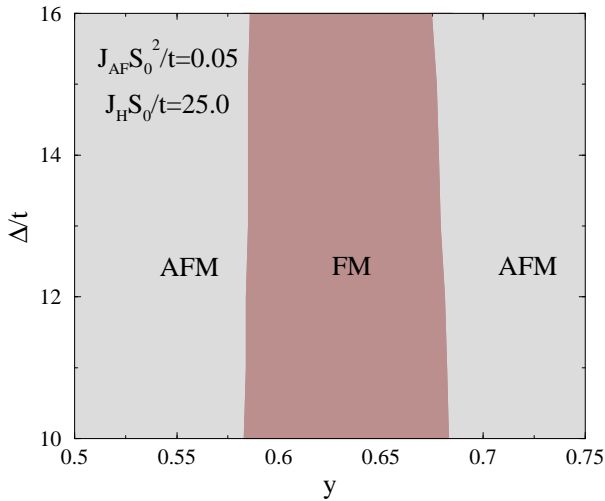}
\caption{The ground state phase diagram  for Fe in a tetrahedral environment
in  the $y-\Delta/t$ plane
at a typical value of $J_HS_0/t=25.0$ and $J_{AF}S_0^2/t=0.05$
}
\label{figure7}
\end{figure}
%%%%%%%%%%%%%%%%%%%%%%%%%%%%%%%%%%%%%%%%%%%%%%%%%%%%%%%%%%%%%%%%%%%%%%%^M
\section{Coulomb Interaction}

The onsite intra- and inter-orbital Coulomb interaction terms given by 
$H_{int}$ (in Eq.\ \ref{H0}) are treated in the mean-field theory. Neglecting
the fluctuation effects, we write 
$ \hat{n}_{i1\sigma}\hat{n}_{i2\sigma^{\prime}}
=<\hat{n}_{1\sigma}>\hat{n}_{i2\sigma^\prime}
+<\hat{n}_{2\sigma^\prime}>\hat{n}_{i1\sigma}-<\hat{n}_{1\sigma}>
<\hat{n}_{2\sigma^\prime}>$, the last term preventing double counting. The averages 
$<\hat{n}_{1\uparrow}>$, $<\hat{n}_{1\downarrow}>$,
$<\hat{n}_{2\uparrow}>$, 
$<\hat{n}_{1\downarrow}>$ are calculated from the eigenvectors
iteratively through successive diagonalization of the Hamiltonian. Self-consistency 
has been achieved when all the averages $<\hat{n}_{i,\sigma,\alpha}>$ and the ground
state energy converge to within 0.01 $\%$ or less. 

It is well known \cite{dagrev,hotta} that in the large J$_H$ limit, the 
Coulomb repulsion $U$ between up and down spin electrons at the same orbital is
ineffective in the mean-field theory. In this limit doubly occupied orbitals are
energetically costly and generally avoided. However, in the case of Fe$^{+2}$, one 
of the $e_g$ orbitals has to be doubly occupied. In that case, in the mean-field type
argument, a replacement of J$_H$ by J$_{H}+U$ in the doubly occupied orbitals takes 
care\cite{hotta} of this repulsion. One then absorbs $U$ in the value of J$_H$ 
appropriately and does not consider it explicitly. In the following, therefore, 
we keep the value of $U$ to be zero. There is, however, a strong effect of 
the inter-orbital Coulomb interaction ($U^\prime$) on the phase diagram as 
we discuss in what follows. 

In Fig. 8 we present the effect of inter-orbital Coulomb interaction ($U^\prime$) 
on the ground state magnetic phase diagram for typical values of $J_HS_0/t$ and 
$J_{AF}S_0^2/t$, keeping $\Delta=0$. As $y$ increases, the FM phase appears as
in the previous figures. At large $U^\prime$ and when almost all the Fe ions
are in the +2 valence state, the AFM phase reappears at the right top corner
of Fig. 8. In the presence of inter-orbital Coulomb interaction the energies 
in the high Fe$^{+2}$ region are primarily
dominated by localized exchange interactions. The competing interactions now have 
the energy scales $-\frac{t_{12}^2}{U^\prime}$ and $-\frac{t_{22}^2}{J_H}$. In 
the limit $U^{\prime} > J_H\,\,$ ($J_{H}+U$, if $U$ is considered explicitly), the 
second term would provide extra gain in energy and the AFM phase should stabilize. 
A transition 
from FM$\to$AFM will therefore occur as $U^{\prime}$ exceeds $J_{H}$ for
$y=0.75$. As $y$ reduces from 0.75, a larger $U^\prime$ is required for the
transition leading to the region of AFM at the top right hand corner in 
Fig. 8 as shown.  

%%%%%%%%%%%%%%%%%%%%%%%%%%%%%%%%%%%%%%%%%%%%%%%%%%%%%%%%%%%%%%%%%%%%%%%^M
\begin{figure}
\includegraphics[clip=true,width=80mm]{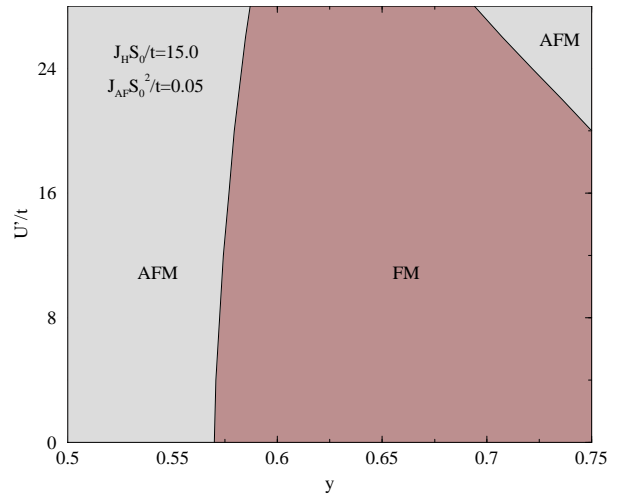}
\caption{Phase diagram of the Hamiltonian for Fe in a tetrahedral environment
showing the effect of $U^\prime$ (see Eq. \ref{H0})  on the 
ferromagnetic and antiferromagnetic phases
}
\label{figure8}
\end{figure}
%%%%%%%%%%%%%%%%%%%%%%%%%%%%%%%%%%%%%%%%%%%%%%%%%%%%%%%%%%%%%%%%%%%%%%%^M

\section{F\lc{e} in octahedral position}

So far we have investigated the case where the doped Fe ions are in  the 
tetrahedral Zn positions of \zn with Fe$^{2+}$ and Fe$^{3+}$ valence states. 
Let us now examine the case when Fe is doped into octahedral Ga positions.
In the octahedral crystal field the energy levels of Fe will be split into 
a triply degenerate set of $t_{2g}$ levels and a doubly degenerate set 
of $e_g$ levels. The $t_{2g}$ levels in this case have lower energy than the
$e_g$ levels, contrary to the tetrahedral case. With this arrangement
of orbitals, the extra (6th.) electron in Fe$^{2+}$ will occupy the $t_{2g}$ 
level. The preliminirary band structure results with 25$\%$ Fe substitution in the Ga
sites \cite{pisani} corroborate this scenario.
The overlap integrals between the $t_{2g}$ orbitals are calculated as
usual from the Slater-Koster integrals \cite {slater}. In this scenario, 
we consider the $t_{2g}$ electron as itinerant for reasons similar to the 
ones discussed in the case of Fe in tetrahedral position. We, therefore, 
use the overlap integrals between the $t_{2g}$ orbitals in the kinetic 
energy term of Hamiltonian Eq. \ref{H1}. We note at this point that in the
spinel crystal structure of \zn these octahedral centers, occupied
by the Ga atoms, are arranged in a tetrahedral fashion among themselves 
and hence are geometrically frustrated. We have not considered
this geometrical frustration in the present calculation because the
experimentally observed ratio of Fe to Ga is fairly small (1:6). The Fe 
atoms are assumed to be arranged in a cubic environment for the present 
calculation and hence there is no frustration.  

In the kinetic energy term of the Hamiltonian given by Eq. \ref{H0} the orbital
indices $\alpha$ and $\beta$ now take the values 1,2 and 3 which 
represent $xy$,$yz$ and $zx$ orbitals respectively. This Hamiltonian will reduce in 
{\bf k}-space to the form  
of Eq. \ref{H1} as in the case of tetrahedral doping except from the fact that now  
$\epsilon_{\bf k}^{\alpha\beta}$ is a 3x3 matrix with the elements 
given by 
\begin{eqnarray}
\epsilon_k^{11}&=&-2t^{\prime}(cosk_x+cosk_y)\nonumber\\
\epsilon_k^{22}&=&-2t^{\prime}(cosk_y+cosk_z)\nonumber\\
\epsilon_k^{33}&=&-2t^{\prime}(cosk_x+cosk_z)
\label{eq5}
\end{eqnarray}

Here $t^{\prime}$ is the magnitude of the hopping integral between the
neighbouring $\pi$-bonded $xy$ orbitals in the x,y direction. Note that the 
inter-orbital
overlaps turn out to be zero in this case \cite{slater}. Following similar 
procedures 
outlined in the tetrahedral case we calculate the ground state magnetic 
phase diagram and observe that (as shown below) the ferromagnetic phase is 
stable in a wide 
range of parameter space in this case as well. However, we have not considered 
the Coulomb interactions ($H_{int}$ in Eq. \ref{H0}) in the present calculation 
of the phase diagram.  
%%%%%%%%%%%%%%%%%%%%%%%%%%%%%%%%%%%%%%%%%%%%%%%%%%%%%%%%%%%%%%%%%%%%%%%^M
\begin{figure}
\includegraphics[clip=true,width=80mm]{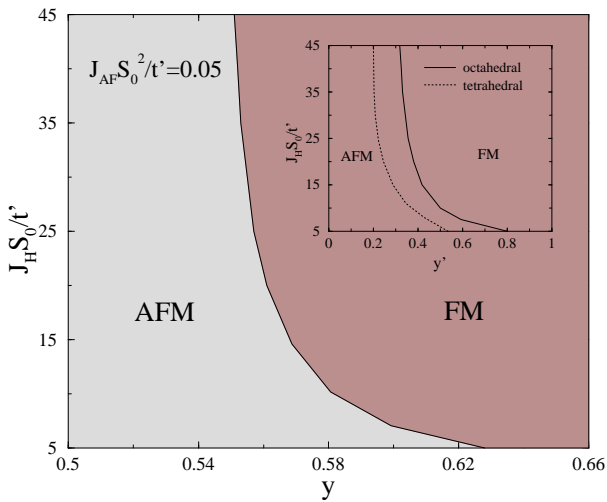}
\caption{Phase diagram for the ground state of the Hamiltonian (with $U=U^\prime=0$)
 for Fe in an octahedral
 environment in the $y-J_{H}S_0/t^\prime$ plane with 
$J_{AF}S_0^2/t^\prime=0.05$. In the inset we show the shift of the AFM-FM phase 
boundary
compared to that of the tetrahedral case (x-axis is rescaled such 
that $y^\prime=0$ corresponds to all Fe$^{3+}$ and $y^\prime=1$ to all 
Fe$^{2+}$ in order to make a comparison)
}
\label{figure9}
\end{figure}
%%%%%%%%%%%%%%%%%%%%%%%%%%%%%%%%%%%%%%%%%%%%%%%%%%%%%%%%%%%%%%%%%%%%%%%^M

In Fig.\ \ref{figure9} we present the ground state phase diagram
in the plane $y-J_{H}S_0/t^\prime$ with a fixed value of 
$J_{AF}S_0^2/t^\prime=0.05$. The three t$_{2g}$ orbitals have been taken
as degenerate (i.e., $\Delta=0$) and $y$ is the electron concentration 
in the $t_{2g}$ levels ranging from $y=1/2$ (corresponding to all Fe$^{3+}$)
to $y=2/3$ (all Fe$^{2+}$, 4 electrons in $t_{2g}$). A comparison
with the earlier phase diagram, for Fe in a tetrahedral position
(Fig.\ \ref{figure5}), is shown in the inset. We observe a shift of
the AFM-FM phase boundary towards higher concentration of Fe$^{2+}$ in the
octahedral case. The absence of off-diagonal hopping among the t$_{2g}$ 
orbitals (Eq. \ref{eq5}) reduces the effective KE gain in the double exchange
mechanism. This, in turn, makes the FM phase less stable compared to the 
tetrahedral case. The antiferromagnetic phase, therefore, stabilizes over a 
wider region in the phase diagram. 

Nevertheless, we still observe a stable ferromagnetic phase in a moderate to 
high range of doping by Fe$^{2+}$ ions. Finite JT splitting
of t$_{2g}$ orbitals may have interesting effects on the stability of the
ferromagnetic phase in the high Fe$^{2+}$ limit. Since the inter-orbital hopping 
is zero, even a small but finite JT splitting may induce a cooperative 
staggered JT distorted order of the Fe$^{2+}$ ions in the ferromagnetic phase 
to maximize the KE. This would make the ferromagnetic phase increasingly 
destabilized against the AFM phase, depending on the value of $\Delta$, as 
well as against defects and other impurities present in the system\cite{ftn2}. 
Coulomb correlations, particularly the inter-orbital Coulomb correlation 
($U^\prime$), is expected to have a  strong effect in stabilizing the AFM phase 
in this case due to the reduced mobility of the carriers in the $t_{2g}$ 
orbitals as argued above. Following the arguments in Sec. V, the AFM phase at 
the right top corner in Fig.\ \ref{figure8} is expected to appear at a lower value of 
$U^\prime$ now.

\section{Discussion}

We observe from the above study of Fe doped \zn that ferromagnetism could 
result from a delicate competition between double exchange, favouring the FM 
spin order, and superexchange, favouring AFM order, when both
Fe$^{2+}$ and Fe$^{3+}$ valence states are present. Though the present
experimental situation in this system does not make the limit of all Fe$^{2+}$ 
in tetrahedral positions a very relevant one,  there is  nevertheless a rich 
underlying physics connected with the interplay between dominant Jahn-Teller, 
double exchange and superexchange interactions in this limit.
Finally, a special note is in order for the case where the doped Fe ions
go to both tetrahedral and octahedral environments replacing Zn and Ga ions
respectively. In this mixed situation, it is necessary to first find out the 
extended and localized states from a careful density functional calculation.
As the $e_g$-$t_{2g}$ orbital overlap is negligible, it could be possible 
that in the mixed situation the double exchange mechanism would operate within 
the Fe ions that belong entirely to one kind of crystalline environment
(i.e., either in the tetrahedral or in the octahedral positions) and hence
only a fraction of the Fe that are doped into the system would take part in
the ferromagnetic long range order \cite{risbud}. However, the model is 
expected to work as long as there are mobile electrons coupled to a relatively 
localized spin background.

We have been discussing in this work the 
 carrier-mediated ferromagnetism which is believed to explain the magnetic properties
of various dilute magnetic semiconductors with available free charge carriers 
\cite{dietl,pearton}.
For relatively localized systems with no free carriers other alternative 
mechanisms such as that of the  bound magnetic polaron model have also been 
proposed 
\cite{berciu,kaminski}. One such example is Ga(Mn)N where it has been
suggested that the Mn ions are in d$^5$ configuration plus a localized hole
and  this localized hole forms a singlet with a Mn d-electron 
(Zhang-Rice polaron) 
which then moves through the Mn sublattice  
and mediates ferromagnetic order\cite{dietl1,sanvito1}.

The system we consider in this work 
 is different from other III-IV or II-VI semiconductors
in some respects. First of all the doping concentration (50$\%$ or 25$\%$) is quite 
high
compared to, for example, the 6$\%$ doping in Ga(Mn)N. Secondly there is a 
possibility of Fe having a mixed valence state of Fe$^{2+}$/Fe$^{3+}$. Thirdly the
existence of two kinds of crystalline environment to which
Fe can be doped is not present in the known DMS semiconductors.

DMS Systems like Ga(Mn)As have indeed shown dependence of T$_c$ on the so called 
antisite defects. Long range order in DMS systems is known to show sensitivity to 
disorder as well. But in the present context, disorder (which we did not consider) 
may not be that crucial for the underlying mechanism of magnetism proposed. 
Unlike the usual DMS materials, the doping is fairly high to be in the 
impurity dominated regime. The disorder may affect the values of exchange interaction 
and electron mobility, thereby shifting the phase boundaries slightly, but the overall topology of the ground state phase diagram will remain unaffected. Therefore, we
believe that at least to a first approximation, effects of disorder could
be neglected.

Finally, concerning the model (Eq. \ref{H0}),
there are also very interesting issues related to it like the 
possibility of phase separation and canted spin structures, possible orbital 
ordering and low dimensional spin orders which have not been investigated here. 

\section{Conclusion} 

In conclusion, based on the band structure results we have presented an effective model
for magnetism in Fe doped  \zn  which predicts a stable 
ferromagnetic phase when both Fe$^{2+}$ and Fe$^{3+}$ valence states are 
present. 
If only Fe$^{3+}$ is present -as reported in the Mossbauer
spectroscopy\cite{risbud}- it is not possible to get ferromagnetism via this
model,  
an insulating AFM state would have been the most likely ground 
state. As the system is not very dilute and the transition temperature
is quite high, it is likely that ferromagnetism in this system is
driven by the kinetic energy of mobile electrons via double exchange 
rather than interaction between localized impurities. A high degree of 
delocalization of the extra electron in Fe$^{2+}$ could also explain 
the observation of only Fe$^{3+}$ states in the Mossbauer experiments.  
More experiments are needed to be done in order to unambiguously detect 
the states of Fe in \zn. A study of the ground states for a range of doping 
concentrations would be very useful. Photoemission experiments backed by 
detailed first principle calculations are also indispensable to delineate 
the relevant orbitals that participate in the double exchange mechanism.
\vspace{0.2cm}

\noindent {\bf Acknowledgements} We would like to thank L. Pisani and 
R. Seshadri for useful discussions.  RV  acknowledges  financial 
support from the German Science Foundation.


\begin{thebibliography}{99}
\bibitem{ftn3} 
Presence of doubly occupied $e_g$ orbitals in both initial and 
final states ensures that the intra-orbital Coulomb repulsion ($U$),
contributes to the total energy in 
both states.
\bibitem{ftn2} 
Such orbitally ordered states may be broken easily by
defects in these disordered systems and long range transfer of charge 
carriers inhibited.
\end{thebibliography}
\end{document}